# THE CHALLENGES AND IMPACT
# OF PRIVACY POLICY COMPREHENSION

*Research paper*

Korunovska, Jana, Vienna University of Economics and Business, Vienna, Austria, jana.korunovska@wu.ac.at

Kamleitner, Bernadette, Vienna University of Economics and Business, Vienna, Austria, bernadette.kamleitner@wu.ac.at

Spiekermann, Sarah, Vienna University of Economics and Business, Vienna, Austria, sarah.spiekermann@wu.ac.at

## Abstract

*The new information and communication technology providers collect increasing amounts of personal data, a lot of which is user generated. Unless use policies are privacy-friendly, this leaves users vulnerable to privacy risks such as exposure through public data visibility or intrusive commercialisation of their data through secondary data use. Due to complex privacy policies, many users of online services unwillingly agree to privacy-intruding practices. To give users more control over their privacy, scholars and regulators have pushed for short, simple, and prominent privacy policies. The premise has been that users will see and comprehend such policies, and then rationally adjust their disclosure behaviour. In this paper, on a use case of social network service site, we show that this premise does not hold. We invited 214 regular Facebook users to join a new fictitious social network. We experimentally manipulated the privacy-friendliness of an unavoidable and simple privacy policy. Half of our participants miscomprehended even this transparent privacy policy. When privacy threats of secondary data use were present, users remembered the policies as more privacy-friendly than they actually were and unwittingly uploaded more data. To mitigate such behavioural pitfalls we present design recommendations to improve the quality of informed consent.*

# 1 Introduction

When thinking about how to ensure user control in digital computing environments we often count on users' rationality. Many scholars and practitioners build digital applications against the assumption that users will consciously absorb and consider the information and choices they are being given online. For example, when online users decide whether to share their personal data (PD), they are expected to take this decision seriously, invest some time in choosing the most trustworthy service provider, and share data only with those who deserve it. However, the everyday practice of users paints a different picture.

For example, Internet users disclose enormous amounts of PD particularly when they engage in Social Network Service (SNS) platforms. By default, the SNSs make these self-disclosures publicly available, which makes users vulnerable to the threat of PD visibility. More importantly, the SNS platforms additionally make use of the PD provided for social networking purposes (primary purpose), for other, secondary purposes. These secondary purposes involve among other things sharing and selling of the PD to third parties (Federal Trade Commission, 2014; Jentzsch, 2017; World Economic Forum, 2012). Thus, disclosure of PD on SNSs also exposes users to the threat of secondary PD use.

While these SNS practices are in line with the privacy policies signed off by users, an abundance of privacy scandals have shown that users' privacy is indeed often at risk (Krasnova, Spiekermann, Koroleva, & Hildebrand, 2010; Xu, Dinev, Smith, & Hart, 2011). As more digital platforms and applications apply business models based on secondary PD use, and as these platforms enter even more aspects of our daily lives, such privacy breaches are expected to rise (Zuboff, 2019).

One key to preventing privacy infringements should be the privacy policies, which ideally would make the users aware of the privacy risks and help them adjust their behaviour. In reality though, this barely happens. Despite the general availability of privacy policies, users have little awareness of what data are collected and what threats this may entail (Acquisti, Brandimarte, & Loewenstein, 2015). A presumed key reason for this lack of awareness is that policies are too lengthy, hidden in the small print, written in legalese language and thus are hard to comprehend (Martin, 2015; McDonald & Cranor, 2008; Milne & Culnan, 2004; Strahilevitz & Kugler, 2016). As a result, users mostly provide "consent" quickly and rather to get rid of the nuisance than actually providing an *informed* decision about the use of their data. This behaviour has become habituated and has not changed since the introduction of the Europe's General Data Protection Regulation (GDPR).Thus, the calls for more comprehensive, prominent, short and simple privacy policies continues (Cate, 2010; Data Ethics Comission, 2019; Milne & Culnan, 2004; Pollach, 2007). The assumption is that better privacy policies will finally lead to *truly* informed consent and that people will become more rational actors as portrayed in academic privacy theories, such as the APCO model (Smith, Dinev, & Xu, 2011). In these academic theories, users rationally adjust their privacy behaviour according to their personal preferences and share more information with privacy-friendly providers, or conversely, cautiously restrict data sharing in privacy-invasive ones. In fact, this hope has driven Article 7 of the GDPR, as well as the recommendations of the Federal Trade Commission (FTC) to prescribe the prominence of privacy policies and the use of clear and plain language (Council of European Union, 2016; Federal Trade Commission, 2012).

However, it is not yet clear whether transparent privacy policies will indeed affect informed consent and rational PD disclosure. First, transparency is an ambiguous concept, not least because it incorporates both simplicity and prominence, which may have different impact on comprehension (Nissenbaum, 2011; Walker, 2016). Second, studies have shown that privacy policies can also nudge and trick users into irrational disclosure especially if they signal trust (Acquisti et al., 2015). That said, to our knowledge, no study to date has tested the effect of policy *transparency* while controlling for user comprehension of the policy and subsequent behaviour. Even if prominent, users may still habitually ignore the policy or simply miscomprehend it (Acquisti, Adjerid, & Brandimarte, 2013; Martin, 2015). Thus, even if companies became transparent and offered their privacy policies in the clearest possible way, we still do not know whether privacy will then be better protected. The aim of this article is to investigate this.

For that purpose, on a use case of a SNS, we present an experiment that tries to answer two fundamental research questions: RQ1) Will transparent privacy policies lead to actual policy comprehension? And RQ2) Will policy comprehension prompt users to rationally align their willingness to join the SNS and more importantly their disclosure behaviour with the degree of privacy-friendliness offered by the SNS? In addition to addressing these two overarching questions, we contribute to the literature by taking a nuanced look at the type of privacy threat present in the policy: RQ3) Will policy comprehension, willingness to join the SNS and disclosure behaviour depend on the specific type of threats present in the privacy policy?

Specifically, we focus on the two most prominent privacy threats in SNSs: 1) the threat of PD visibility (sometimes referred to "accessibility" threat); and 2) threat of secondary PD use (Earp, Antón, Aiman-Smith, & Stufflebeam, 2005; Krasnova, Günther, Spiekermann, & Koroleva, 2009; Steinfeld, 2016). In our experiment, we vary the presence of both privacy threats when manipulating the privacy-friendliness of policies, i.e., we manipulate whether the general public can access and see the personal data of the SNS user and whether the SNSs has the right to use the data beyond the primary SN purpose, for example sell it for commercial gain. Consequently, we see the impact of the various policies on user comprehension, willingness to join the SNS and self-disclosure on the SNS.

This research adds to earlier privacy research in e-commerce settings (Adjerid, Acquisti, Brandimarte, & Loewenstein, 2013; Andrade, Kaltcheva, & Weitz, 2002; Jensen, Potts, & Jensen, 2005; LaRose & Rifon, 2007; Liu, Marchewka, & Ku, 2004; Metzger, 2004). To our knowledge, it is the first study to investigate how different types of transparent privacy threats affect comprehension, willingness to join SNSs and actual disclosure behaviour (not just intentions). Specifically, it looks into the interplay between the content of the privacy policy (different degrees of privacy-friendliness, i.e., privacy-intrusiveness), comprehension of the policy, the willingness to join the SNSs and disclosure on the SNSs.

## 2 Related Work: The Role of Transparent Privacy Policies for Information Privacy

Many theories try to describe and explain self-disclosure online (for reviews see Barth & de Jong, 2017; Gómez-Barroso, 2018; Kokolakis, 2017; Smith et al., 2011). Barth and de Jong (2017) discern three theoretical perspectives on user privacy behaviour: The first assumes rational users who decide whether to disclose based on a "privacy-calculus". Privacy-calculus refers to the decision-making process that happens before users reveal something online. Here users weigh on the one hand, the privacy risks of disclosing PD, the trustworthiness of a platform etc., and on the other hand, the benefits they would receive if they disclosed the PD. If the calculated benefits and the trust outweigh the privacy risks then users are expected to disclose PD (Culnan & Armstrong, 1999; Dinev & Hart, 2006). Most prominent in this respect is the APCO model, which assumes a relationship between users' awareness of the privacy practices and self-disclosure (Smith et al., 2011). The second perspective also assumes a 'privacy calculus' but does not find users to be as rational as expected. Disclosure decisions are found to be bounded by decision-making biases and cognitive limitations. Users are found to make systematic errors when disclosing PD, unwillingly putting their own privacy at risk (Acquisti et al., 2015). The third approach assumes that users do not engage in any rational privacy calculus and simply fail to account for privacy risks altogether (Quinn, 2016).

Which of these scientific perspectives best reflects reality holds fundamental implications for the design of user-centric privacy protection. Depending on perspectives, users are either able to self-protect when provided with all relevant information in a user-friendly way or they need protection through proactive privacy by design that is present by default and is enforced by regulators (as called for by the Data Ethics Commission, 2019). Despite a wealth of research, the empirical evidence on which of the three perspectives best reflects reality is not straightforward and there are only few experimental insights (Dinev, McConnell, & Smith, 2015; Gómez-Barroso, 2018; Smith et al., 2011). Until recently, the majority have followed the rational user approach and it has been an exception "for privacy researchers to consider alternative models and explanations outside the APCO model" (Dinev et al., 2015, p. 640). For rational users to make informed decisions, policies should therefore provide all necessary information for a privacy calculus.

That said, even ignorance can be rational when the cost of getting the information outweighs the benefit of disclosure (Flender & Müller, 2012). For example, McDonald and Cranor (2008) estimated that users would need to spend 19 days if they only skimmed every privacy policy they encounter online in a year. Thus, users practically never read these hard to understand texts (Milne & Culnan, 2004; Steinfeld, 2016). Following this practical reality, uninformed PD disclosure is assumed to be the consequence of policies' inadequacy as an information tool (Earp et al., 2005; Milne, Culnan, & Greene, 2006; Pollach, 2007; Strahilevitz & Kugler, 2016). Consequently, designers, scholars and policy makers have called for policies to become more transparent so that they can be easily comprehensible and absorbed in little time (Data Ethics Comission, 2019). This means they need to be simpler, shorter and more prominent.

## 2.1 Effects of transparency on policy comprehension

It is plausible to assume that the majority of users will read and comprehend a transparent, i.e., short, simple and prominent privacy policy. An eye-tracking study in fact found that users carefully read a privacy policy when it was presented prominently but at best skimmed it when they had to click on a link in order to access it (Steinfeld, 2016). Since this is also what most of policy makers and proponents of the rational user approach assume, and in line with our first research question, we hypothesize:

**H1:** The majority of SNS users will comprehend a transparent privacy policy

However, we also challenge this seemingly obvious hypothesis. This is because when Internet users have limited time or attention due to distraction, they may in fact *not* fully comprehend privacy policies (Acquisti et al., 2013; Martin, 2015). As already mentioned, transparency incorporates two aspects: simplicity and prominence (Nissenbaum, 2011; Walker, 2016). While simplicity is likely to increase user comprehension (Jensen et al., 2005; Pan & Zinkhan, 2006), prominence can also have negative effects, because it might falsely assure users that there is nothing to hide (Acquisti et al., 2013; Walker, 2016). Even if users read the policies, apparent transparency can lull comprehension. Ample evidence suggests that people see what they expect to see even if what they expect to see is not there (Janiszewski, 2008). Consequently, users "project their privacy expectations onto the privacy notice" (Martin, 2015, p. 219). So if users interpret transparency as a signal that there is nothing to hide and nothing to worry about, then they may miss, misperceive or miscomprehend threats to their privacy even after a cursory reading of privacy policies. We coin this behaviour as a *miscomprehension in a privacy-unsafe direction*. In contrast, users may also interpret prominent privacy policies as a privacy warning and become more concerned and overly cautious even if there are no real privacy threats (John, Acquisti, & Loewenstein, 2011). This is what we call *miscomprehension in a privacy-safe direction*. In sum, transparency can both foster and hinder a comprehension-based (informed) consent. We will therefore test here for the privacy-safe and privacy-unsafe directions of users' *mis*comprehension, if such miscomprehension is evident.

## 2.2 Comprehension impact: The effects of transparent privacy-friendliness

To our knowledge, no study to date investigated the interplay between varied degrees of privacy-friendliness and transparency in the context of SNS platforms. Related evidence shows that users disclose more PD when they see transparent privacy-friendly policies or symbols and disclose less when they see privacy-invasive policies or privacy warnings (Andrade et al., 2002; Gideon, Cranor, Egelman, & Acquisti, 2006; Jensen et al., 2005; LaRose & Rifon, 2007; Tsai, Egelman, Cranor, & Acquisti, 2007; Wirtz, Lwin, & Williams, 2007). Yet, this only seems to hold true for users who really pay attention to the policy and, again, correctly comprehend it. In an e-commerce study by Metzger (2004) for example, users disclosed the same amount of PD regardless of whether the policy was simple and privacy-friendly, vague and friendly, or entirely lacking. Notably, about a third of participants in the study failed to notice the prominently displayed policies. Further two studies have shown that the content of privacy policies matters only when participants clicked on them (Jensen et al., 2005) or were undistracted during the policy presentation, constraining the effect of its content to ideal circumstances (Acquisti et al., 2013).

Against this background, we do not expect transparency alone to increase users' interest or lead all users to adjust their PD disclosure in accordance with the privacy-friendliness of the policy. We only expect this from users who actually comprehend the policies. Thus, related to RQ2, we hypothesize:

> **H2a:** Policy comprehension moderates the relationship between the privacy-friendliness of a policy and the willingness to join the SNSs.

> **H2b:** Policy comprehension moderates the relationship between the privacy-friendliness of a policy and self-disclosure.

## 2.3 The effects of specific privacy threats on comprehension, willingness to join SNS and personal data disclosure

Thus far, we have discussed privacy-friendliness as a unidimensional concept, with a policy being either privacy-friendly or privacy-invasive. However, privacy policies in reality contain distinct elements that can be more or less privacy-friendly. For example, Facebook may be privacy-friendly when it comes to their users' visibility, which they can self-configure by now. But at the same time the company seems rather privacy-unfriendly when it to comes to their business model of selling their users' data for secondary purposes to third parties. To date and to our knowledge, there is little evidence on how consumers react to such nuanced threats within one policy. A majority of studies have used scales that combine general privacy concerns (e.g., "How concerned are you about your personal privacy on this website") with concerns about specific privacy threats (e.g., "How concerned are you about this website sharing your personal information with other parties?" or "Are you concerned that an email you send may be read by someone else besides the person you sent it to?") (Buchanan, Paine, Joinson, & Reips, 2007; Kuppelwieser & Sarstedt, 2014; Metzger, 2004; Smith, Milberg, & Burke, 1996). Yet, not all of such individual threats addressed in a policy may be deemed relevant by users (Earp et al., 2005; Krasnova et al., 2009; Milne et al., 2006; Van Slyke, Shim, Johnson, & Jiang, 2006). For example, users were shown to look significantly longer on the sections "information we collect", "how we use your personal information" and "sharing and disclosure of personal information" than on other subject matters (Steinfeld, 2016). In other words, specific privacy threats, rather than an overall level of privacy-friendliness may drive users' disclosure decisions.

For this reason, we have discerned two of the most prevalent and dangerous privacy threats in the context of SNSs presented in the Facebook example: (1) the threat of PD visibility (for e.g. to an unwanted or unknown audience) and (2) the threat of secondary PD use (for e.g. selling the PD to third parties). Whereas most regulatory privacy discussions focus on the threat of secondary PD use, the threat of PD visibility has been a large concern of SNSs users themselves (Krasnova et al., 2009; Raynes-Goldie, 2010; Spiekermann & Korunovska, 2017).

Users have reason to be concerned about both of these threats. Yet, we propose that they will not react to them equally strongly. Specifically, we propose that users react more strongly to the more easily comprehensible threat of data visibility than to secondary data use. The reason we assume stronger reactions to visibility is twofold: First, visibility is a dominant theme in the privacy discourse specific to SNS. SNS providers themselves foster a focus on visibility when they talk about privacy. For example, under the privacy settings on Facebook the three options are "Who can see my stuff?", "Who can contact me?", and "Who can look me up?" All of these are visibility concerns. Likewise, seven out of eight options in the privacy settings on Twitter are related to PD visibility while only one is related to secondary PD use.

Second, an argument has been made that users need to see or at least be able to imagine their digital blueprint (i.e. as a personal asset they own) in order to comprehend and protect it (Kamleitner &

Mitchell, 2018). While people know about the power imbalances that can result from knowing more or too much about another person, they cannot imagine the nature of their self-generated Big Data and the impact these data have in today's PD markets where PD is put to secondary uses. Simply, users are unable to imagine "surveillance capitalism" because of its *unprecedentedness* in history (Zuboff, 2019).

Against this background, we argue – against better knowledge of the experts – that the threat of PD *visibility* will be more salient and more important to SNS users. As a result, SNS users will be more likely to comprehend and react to threats of data visibility than to threats of secondary data use. Therefore, for our third research question, we hypothesize:

**H3a:** SNS users are more likely to comprehend transparent privacy policies that entail threats of PD visibility than threats of secondary PD use

**H3b:** Conditional on policy comprehension, the willingness to join the SNS will be higher when there is NO threat to PD visibility than when there is NO threat of secondary PD use.

**H3c:** Conditional on policy comprehension, self-disclose will be higher when there is NO threat to PD visibility than when there is NO threat of secondary PD use.

## 3 Method

### 3.1 Stimulus material

Following FTC and GDPR guidelines on transparency, we designed a privacy policy in a simple tabular and additional textual format argued to best communicate privacy settings (Leon et al., 2012). We focused only on the threats of PD visibility and secondary PD use, which we manipulated as discussed below. Figure 1 depicts two (out of the four) versions of this policy.

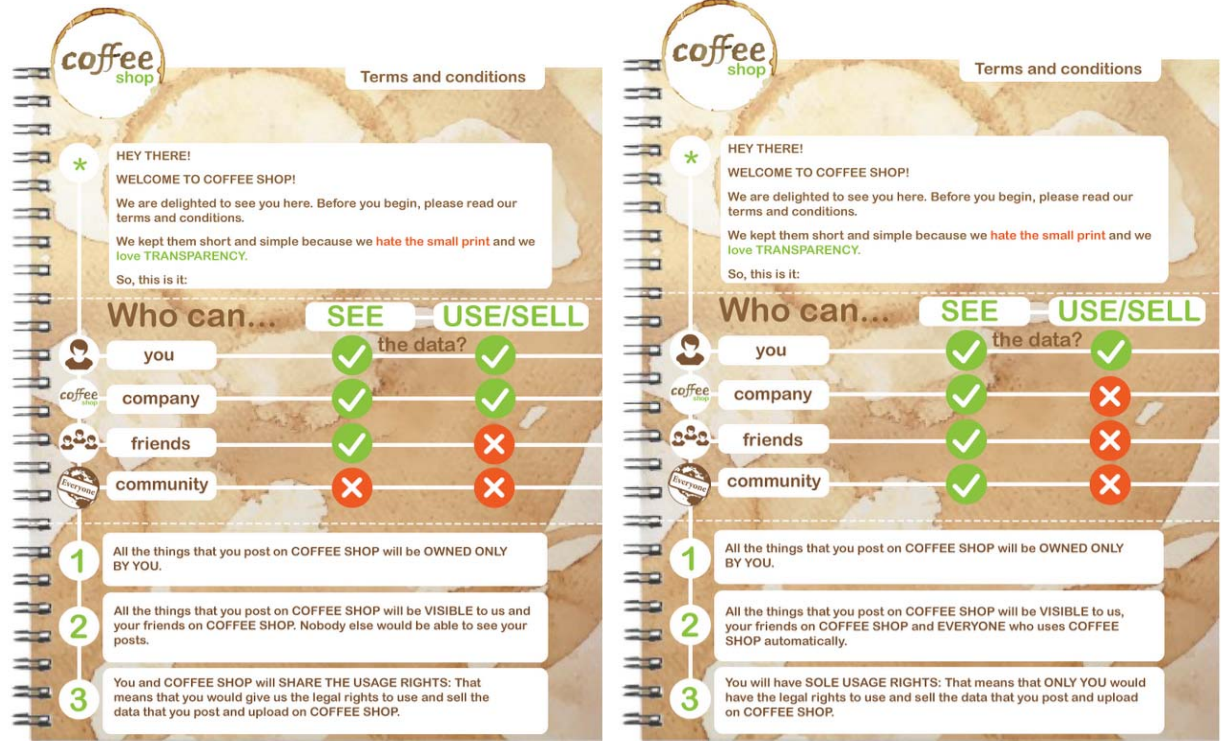

*Figure 1.	Example of a privacy policy on Coffee Shop (version 2 and version 3)*

The policy informs who owns, who can see (visibility information) and who can use/sell PD (secondary data use information) on Coffee Shop, a hypothetical SNS (see below). We pre-tested the perceived transparency of this policy (n=60) and confirmed that this format was perceived as highly transparent in terms of simplicity, clarity and easiness of comprehension. On a eleven-point semantic differential scale with bipolar adjective pairs ranging from -5 to + 5 participants rated the language as clear (M=3.47, SD=2.06, t(60)=6.8, p<.001) and plain (M=3.27, SD=2.04, t(60)=7.6, p<.001), and confirmed that the policy was easy to understand (M= 2.59, SD=2.64, t(60)=7.3, p<.001).

## 3.2 Research design: procedure and measures

We invited respondents to participate in our study by asking them to "help us learn what makes a good social network". In an initial screening section that goes beyond this study, we asked respondents about their attitudes and behaviour on Facebook as well as their basic demographic data. Next, all respondents landed on an information page with the cover design of several news headers. The main headline read "Coffee Shop trumps Facebook. New social network reaches 2 billion users." We asked respondents to imagine that the SNS "Coffee Shop" had emerged on the market, was already overtaking Facebook in popularity and most of their friends had already joined it. We included this information to preclude Facebook's network effect, a known obstacle to user mobility across SNSs. Additionally, participants learned that this new social network has "a very clear, simple, and TRANSPARENT PRIVACY POLICY". We explicitly positioned Coffee Shop as competing against Facebook on the argument of more transparency about privacy, because this is a known concern on Facebook. Respondents clicked next to see what Coffee Shop looked like. They landed on a mock up webpage designed to look like a welcome page where we presented the actual privacy policy of Coffee Shop (Figure 1). The ensured prominence of the privacy policy simulated a situation where policies are unavoidable, following FTC and GDPR guidelines on transparency (see Figure 2 for the study flow).

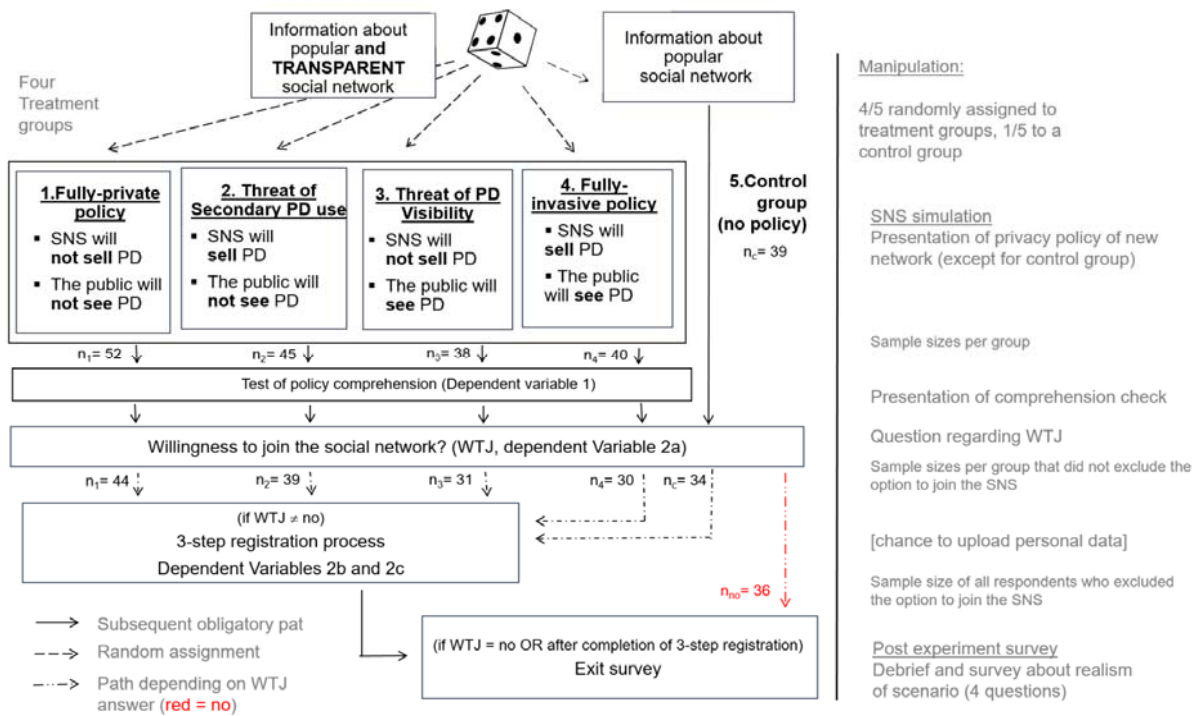

*Figure 2.   Research design and study flow*

We randomly assigned respondents to one of five conditions consisting of a control group and four different privacy policies. The policies differed in terms of the two privacy threats (of PD visibility and secondary PD use) which we varied in a 2×2 between-subject design (Figures 1 and 2). Thus, the first policy was fully privacy-friendly, the forth one fully privacy-invasive, whereas the second and third policies had only one privacy threat each (Figures 1 and 2). We kept all other aspects of the privacy policy constant across conditions. Participants in the control group were asked to join Coffee Shop without any reference to transparency and saw no privacy policy. This group, thus, serves as a benchmark control for what happens when privacy policies are hidden and users are not primed for privacy.

### 3.2.1. Measure of comprehension

In keeping with the cover story, the social network asked their potential new users to demonstrate that they had read and understood the privacy policy ("It really matters to us that before you sign in you really understand your rights at our shop…"). Therefore, immediately after viewing the privacy policy page all experimental groups saw the same table from the previous page and were asked to complete the policy boxes themselves on who can see and use/sell their data on Coffee Shop. If unsure, participants had the opportunity to go back and see the actual policy again before replicating the actual ticks. The correspondence between the ticks made and the actual privacy policy viewed served as a measure of participants' policy *comprehension*, i.e., our first dependant variable. Not having been exposed to a privacy policy, the control group faced no such comprehension test. Since the content of the policies was so simple (see Figure 1), we assume that a fully recalled policy would signal a good level of comprehension that combines an adequate level of subject matter understanding and some working memory awareness of the given privacy threats.

### 3.2.2. Measures for willingness to join and self-disclosure

After the privacy policy comprehension test, respondents indicated how likely it would be that they would want to open an account on Coffee Shop (using a 5-point Likert scale ranging from 1="certainly not" to 5="certainly yes"). All participants who indicated that they had no intention to join at all were thanked and not guided through Coffee Shop any further (n=36). All respondents who indicated that they would be a least somewhat interested in joining, landed on a three-step registration process for the SNS (Figure 2). Here they were asked to provide a name and to write in the "about me" section (step one), to upload a profile picture (step two), and to write a first status (step three). This process mirrored a typical registration process to a SNS. To simulate the potential transfer from Facebook and to comply with the data portability principle in the GDPR, respondents could also click to indicate that they would want to now transfer their name and profile picture from Facebook to Coffee Shop.

This registration stage provided us *with three dependent variables*. First, we used the *willingness to join* the SNS item as a proxy measure of *disclosure intent*. Second, we got two actual self-disclosure variables, namely, the *number of uploaded PD* and the *richness of the self-disclosures*. For the *number of uploaded PD*, we simply summed up the disclosure decisions across the three steps to create the variable. We coded every disclosed item (name, about me, photo, and status) with one. Thus, the resulting *number of uploaded PD* variable ranged between zero and four information items provided. Finally, we assessed the *richness of the self-disclosures* by coding the disclosed content of the status posts with zero if the respondents did not write a status, with one if they wrote only "Hi", "Hello" or similar, and with two if they wrote more personal statuses, thus making richer disclosures.

After the registration process, respondents were redirected to the end page where they were debriefed and asked how realistic they found the SNS transfer scenario to be.

### 3.3 Sample

A professional market research company administered the study to a randomly drawn sample from their online panel of over 30.000 nationally representative Internet users in Austria. 587 panel users opened our survey but 99 did not proceed with the survey after the first page. Further 143 users (29.3% of the remaining sample) were excluded because they did not have a Facebook account and could thus not believably simulate a transfer to a new account. In order to identify respondents who were just clicking through the survey without paying attention, we incorporated two attention-testing questions, one before and one after the experimental manipulation (e.g., please click "rarely" if you are still paying attention).

Nine percent, or 31 out of the remaining 345 respondents did not pass the first attention check, 57 respondents did not pass the second attention check and further 43 respondents dropped out of the survey themselves, yielding a final completion rate of 64%. The final sample consisted of 214 Facebook users (55% female, mean age=40 years; SD=4.5 years) who completed all questions and passed the attention checks. The sample size allowed us to detect medium effects with the recommended statistical power of $1-\beta>.80$ for simple one-way analysis of variance (ANOVA) with five groups, as well as univariate ANOVAs with 5 factors, including interactions (Cohen, 1988; Faul, 2009).

## 4 Results

The experimental groups were structurally equivalent in terms of their demographic makeup, which was representative for a typical Austrian citizen. Namely, all groups were of similar age (F(4,208)=0.52, n.s.), gender (F(4,207)=0.32, n.s.) and had similar levels of education ($\chi^2(1)$=0.72, n.s). Also, all experimental groups found the scenario to be equally realistic (M=3.09, SD=1.23, F(3,209)=1.28, n.s.). Presumably due to the lack of a clear unique selling point of Coffee Shop over Facebook (i.e. transparency), the control group found the transfer scenario to be less realistic than the experimental groups (M = 2.50, SD = 1.26, t(176)=2.51, p<.05).

### 4.1 Effects of transparency on policy comprehension

To test our first hypothesis on whether a majority of SNS users will comprehend a transparent privacy policy, we checked whether the respondents correctly replicated the privacy policies. We compared the simple and unavoidable policies that the participants had seen to those they recalled only seconds later and were surprised by the considerable miscomprehension. Overall, only 53 % of the respondents were able to correctly recall *all* their respective policy's content. This does not statistically differ to half of the respondents (t(174)=.42, p=n.s.). Thus, the result contradicts Hypothesis 1.

To check whether comprehension would differ based on the two specific privacy threats (H3a), we ran a t-test between the users who saw only threats of PD visibility and those who saw only threats of secondary PD use (Table 1, groups 2 and 3). The differences in comprehension between the groups were significant at the α=.10 level, with detectable effect size (t(81)=1.63, p<.10, d=.36). This lends weak support to Hypothesis 3a: transparent threats to PD visibility are slightly better comprehended than threats to secondary PD use.

| Policy | Experimental manipulation | Comprehended: | N | 95% CI | N total |
|---|---|---|---|---|---|
| 1 | Fully privacy-friendly policy (no threats) | 54% | 28 | [40%, 68%] | 52 |
| 2 | Only threats of secondary PD use present | 40% | 18 | [25%, 55%] | 45 |
| 3 | Only threats of PD visibility present | 58% | 22 | [41%, 74%] | 38 |
| 4 | Fully privacy-invasive policy (both threats present) | 60% | 24 | [44%, 75%] | 40 |
| 1-4 | All experimental groups | 53% | 92 | [46%, 71%] | 175 |

Table 1. *Percent) of respondents who comprehended the privacy policy, including confidence intervals (CI) and experimental group sample sizes (N)*

A post-hoc test inquiring about the direction of the policy miscomprehension revealed that participants were most likely to miscomprehend a policy if they were presented with a threat of secondary PD use. Every third respondent (35%) of those who saw a threat of secondary PD use clicked that the SNS will NOT sell their data when in fact it transparently showed it will. As a comparison, 7% of the respondents who were shown NO threat of secondary PD use clicked that the SNS will sell their data. This is a significant difference indicating a systematic error, i.e., more people erring in the privacy-unsafe direction ($\chi^2(2)$=13.8, p < .01). About one in four participants also erred about whether the public can see their data, but this went in the privacy-safe (28%) as well as in the privacy-unsafe direction (24%).

In sum, the majority of the respondents did not comprehend the transparent policies and there was a tendency for users who were shown threats of secondary PD to miscomprehend those threats more often.

Furthermore, the majority of those who miscomprehended the threats of secondary PD use did so in the privacy-unsafe direction.

## 4.2 Comprehension impact: The effects of transparent privacy-friendliness

*Willingness to join the SNS.* To answer H2a, i.e., check whether policy comprehension moderates the relationship between the privacy-friendliness of a policy and the willingness to join the SNSs, we ran two one-way ANOVAs, one for those who comprehended the policies and one for those who did not. The ANOVAs compared the mean willingness to join the new SNS across all experimental groups and yielded a significant result with a medium effect size only when comprehension was considered (Figure 3 left, upper line; $F(3,88)=5.23$, $p<.01$, $f=.39$). A post-hoc test revealed that users who saw and comprehended a fully privacy-friendly policy had higher willingness to join compared to users who saw and comprehended a fully privacy-invasive policy. This was not true for users who miscomprehended the policies. Thus, comprehension leads to rational willingness to join SNSs: The less privacy-friendly a policy is, the less likely a well-informed user is to join a SNS. These results support Hypotheses 2a.

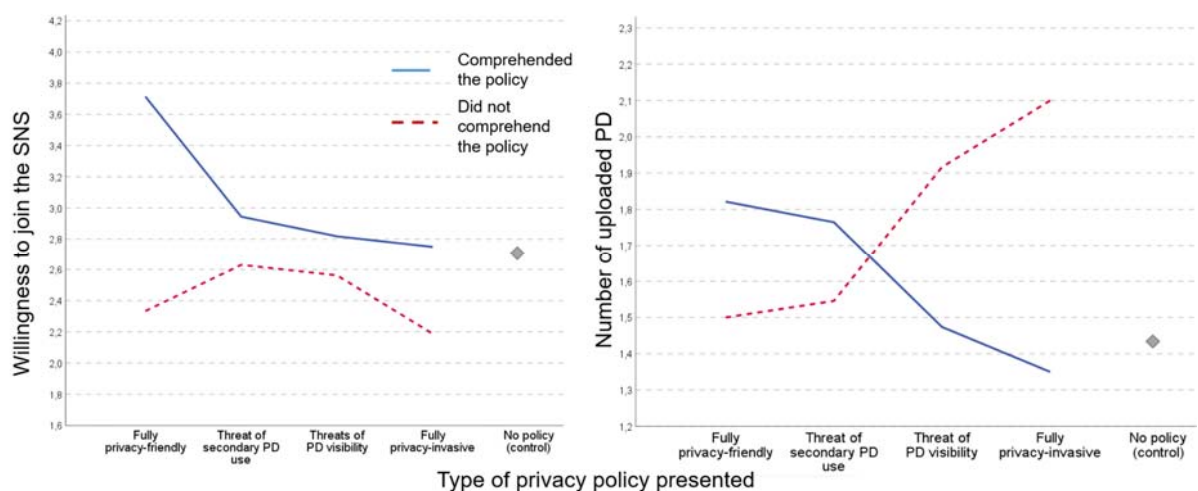

*Figure 3.  Willingness to join the SNS (left) and number of uploaded PD (right) between users shown four different transparent privacy policies, conditional on comprehension*

*Number of uploaded PD.* To test H2b, we used the same method, only with the number of uploaded PD as a dependent variable. The ANOVA that compared the mean number of uploaded PD across all the groups who comprehended the policies yielded a statistically non-significant difference, but with a detectable small effect size (Figure 3 right, upper line, $F(3,80) =.87$, $p=ns.$, $f=.17$). In line with H2b, the post hoc tests revealed a tendency (statistical significance at the $\alpha=.10$ level and notable effect size) for the users who comprehended a fully privacy-friendly policy to upload more PD in the registration process than users who comprehended a fully privacy-invasive policy ($t(46)=1.36$, $p<.10$, $d=.39$). The trend was reversed for those who miscomprehended the policies (Figure 3 right, upper line, $F(3,56) =.64$, $p=ns.$, $f=.18$). This lends weak support for H2b for the number of uploaded PD.

*Richness of the self-disclosures.* To further test H2b, i.e., how comprehension affects the relationship between transparent privacy-friendliness and the richness of the self-disclosure, we further ran two $Chi^2$ tests, one for those who comprehended and one for those who did not comprehend the policies. In line with H2b those who saw and comprehended a transparent privacy-friendly policy disclosed richer PD than those who saw and comprehended a transparent privacy-invasive policy (Table 2, groups 1 and 4, $\chi^2(1)=4.09$, $p<0.05$). This was not evident among users who miscomprehended the policies. These results support H2b for richness of the disclosed PD.

In sum, comprehended privacy-friendliness significantly increased the willingness of users to join the SNS, when compared to comprehended privacy-invasiveness. However, we could only detect a weak increase in the actual number of self-disclosures when users comprehended privacy-friendly policies compared to comprehended privacy-invasive policies. Still, even if not in much higher numbers, those self-disclosures were richer on the SNS with privacy-friendly policies.

| Group | Richness of the self-disclosures: | | | N |
|---|---|---|---|---|
| | Did not write a status | Wrote just Hi, Hello or similar | Wrote a more personal status | |
| 1.Fully private policy (no threats) | 21% | 25% | **54%** | 28 |
| 2.Only threats of secondary PD use (SPDU) | 29% | 29% | 41% | 17 |
| 3.Only threats of PD visibility (PDV) | 42% | 26% | 32% | 19 |
| 4.Fully invasive policy (both threats present) | **50%** | 20% | 30% | 20 |
| Did not see a policy (control group) | 32% | 29% | 38% | 34 |
| Did not comprehend policy | 42% | 36% | 32% | 60 |
| *Note: Groups 1 to 4 are ONLY of users who comprehended the policies* | | | | |

Table 2. Richness *of the self-disclosures: Percent of users with rich, average or no self-disclosures based on privacy-friendliness of policy* and *policy comprehension*

### 4.3 The effects of specific privacy threats on comprehension, willingness to join SNS and personal data disclosure

*Willingness to join the SNS*. To test H3b for the impact of the specific privacy threats, we ran a three-way ANOVA with users' willingness to join the new SNS as a dependent variable and the two aggregated threats as separate factors (independent variables), while controlling for policy comprehension. The results revealed a significant main effect for the presence of a threat to PD visibility ($F(1,167)=3.98$, $p<.05$, $\eta^2=.02$); a significant main effect of policy comprehension ($F(1,167)=14.82$, $p<.001$, $\eta^2=.08$); and a *significant three-way interaction* between all three factors ($F(1,167)=4.42$, $p<.05$, $\eta^2=.03$). This shows that the willingness to join was higher among those who comprehended the policy in general, all those who did not face a threat to PD visibility, as well as all those who did not face any threat and in addition comprehended the policy (Figure 3 left). In other words, if threats of PD visibility were correctly comprehended as present, participants were less likely to want to join the SNS. The presence of secondary data use threats also affected joining intentions, but to a lesser extent, and only when there was no added threat of PD visibility. These results lend support for H3b.

*Number of uploaded PD*. To test H3c for the number of uploaded PD, we ran a three-way ANOVA identical to the one described above, but with the number of uploaded PD as a dependent variable. The ANOVA showed only a significant *two-way interaction* between threats of PD visibility and policy comprehension ($F(1, 136)=4.19$, $p<.05$). The result indicates that the policy only prompted increased amount of disclosed data points if the participants correctly comprehended NO threats of PD visibility, whereas the threat of secondary PD use made no difference. These results lend support for H3b for number of uploaded PD.

*Richness of the self-disclosures*. To test H3c for the richness of the self-disclosures we ran two ordinal regressions, one for those who miscomprehended and one for those who comprehended the policies. We regressed the richness self-disclosure on the type of threats present and again found support for H3b. Among those who miscomprehended the policy, neither threat influenced the richness of the disclosed status. Among those who comprehended the policy, we confirm earlier results. Namely, a SNS privacy policy that entailed a threat of PD visibility led to less personal statuses than a policy without visibility threats ($\beta=-.83$, $p<.05$). In contrast, the comprehension of threats of secondary PD use did not significantly affect richness of disclosure ($\beta=-.35$, n.s).

## 5     Discussion

In the present study, we aimed to answer three research questions. First, will transparent privacy policies lead to actual policy comprehension? Second, will transparency and policy comprehension prompt users to adjust their willingness to join online platforms and disclose more PD on privacy-friendly platforms rather than privacy-invasive ones? And third, will policy comprehension, willingness to join the platform and disclosure behaviour depend on the specific type of privacy threats, specifically threats of data visibility and secondary data use as signalled in the privacy policy?

We find that despite placing the policy prominently and even unavoidably, despite focusing on threats we know users care about, despite using a simple policy design and forcing participants to engage with the policy by asking them to replicate it, half of our sample still did not correctly comprehend the policy. This is even more remarkable if we consider that the sample consisted only of participants who had passed two attention checks, and that respondents could go back and double check the policy. In sum, our first finding suggest that *transparent policies in terms of prominence and simplicity cannot guarantee user comprehension.*

In general, miscomprehension tended to be highest for policies that contained threats of secondary PD use and in favour of the SNS rather than the user. The most frequent recall error was that respondents thought that the company would *not* allow secondary data use, even though it transparently announced that it will. It could be that common practices by popular SNS providers, which draw attention to visibility rather than secondary PD use, have habituated users to disregard these threats (Acquisti et al., 2013; The Internet Society, 2012). Alternatively, transparency signals might have led to unwarranted trust and made users misperceive the policy in line with their trusting expectations (Martin, 2015). It may also just be wishful thinking and undue trust on the side of users. More research is needed to unravel the full reasons behind users' miscomprehension and whether our results are applicable to other digital platforms other than SNSs.

The second key finding is that comprehension does prompt users to rationally react to the privacy-friendliness of the policies. The self-disclosures are furthermore richer on privacy-friendly SNS than on privacy-invasive ones. Nevertheless, the threat of PD visibility has a significantly stronger impact on users than the threat of secondary PD use. Users were not only more likely to comprehend the threat of PD visibility but also once they were aware that their data would *not* be publicly visible, they had higher intent to join the SNS, and disclosed more and richer PD. In contrast, threats of secondary data use only affected the willingness to join under condition that threats of PD visibility were not present. Moreover, the threats of secondary data use did not affect actual disclosure or richness of disclosure when controlling for the threats of PD visibility. This is noteworthy and worrying because secondary data use may be an even bigger actual threat to users in times of surveillance capitalism (Zuboff, 2019). Secondary PD use may pertain to all data, including PD that is harder for users to grasp and for which they are less conscious, such as log-in data etc. As users learn that threats of PD visibility may entail threats of secondary PD use, as was the case with the Facebook-Cambridge Analytica data scandal, and as they gain more right to control secondary data use through the GDPR, it will be interesting to see if they will react more strongly to threats of secondary PD use in the future.

In sum, a large part of our sample of SNS users was unable to provide an informed and hence real consent and – to a large extent – did not disclose data as a function of the risks the policy entails. Users seem so accustomed to all-or-nothing offers in exchange for their data that little energy for caring seems to be left (Borgesius, 2015). And if they care, they do not care for that which poses the greatest threat to their freedom and to democracy at large (that is the secondary use of the data). This puts into question whether the continued call for better policies, such as layered policies, will lead to the political goal of people being better protected. Our results suggest that the rational user hypothesis is at best only valid under specific circumstances (users correctly comprehend the policy) and only for specific privacy threats. Therefore, the current expert call for identifying some regulatory redlines for PD trade in general (e.g. for sensitive data, for child data) or redlines for PD use in certain industries (e.g. scoring) and to pre-define what is allowed and legitimate and what isn't, is highly needed (Data Ethics Commission, 2019). Clear regulation on the limits of secondary PD use is important regardless of whether user consent is provided or not, because our study demonstrates how likely it is that consent is not truly informed. Even a pre-tested highly transparent, simple and accessible policy is likely not to be comprehended. Thus, at minimum, comprehension checks such as proof of direct user comprehension in a captcha-like manner seem worthwhile testing. However, our results show that stricter regulation as discussed above coupled with technological solutions that reduce users' engagement are a better option.

In the past 20 years privacy-scholars have continued to envision intelligent user-centric software agents for privacy that could become part of users' personal data management applications (Kirrane & Decker, 2018; Langheinrich, 2005). These come with privacy-friendly defaults (set by experts), help building user awareness for various privacy threats, and automatically engage with the machine-readable privacy policies of data collectors on behalf of their users (Cranor, Dobbs, Egelman, Hogben, Humphrey, & Schunter, 2006; Kirrane & Decker, 2018). Since such tools could automate many of today's cumbersome one-to-one privacy negotiations with each online player (e.g. reading notices, even if they are simple), our results show that this kind of semi-automated privacy-agent is needed more than ever.

Our results also point to an immediate need for data collectors and IS designers to highlight the specific threat of secondary data use. This could be done with the help of visualization of these practices (Angulo, Fischer-Hübner, Pulls, & Wästlund, 2015). In addition, enhanced user control over secondary data use might help mitigate the current failure of self-protection. Though most SNSs allow users to easily restrict data visibility (Hartzog, 2010) they do not allow similar user-friendly controls for secondary data use. Potentially, policy makers will need to force services to offer similar controls for secondary data use without restricting the availability of the service itself and/or demand privacy by design and by default; two further recommendations equally made by the Data Ethics Commission (2019).

## 5.1 Limitations

Our sample was representative only of Internet users in Austria. Since privacy norms and expectations are culturally dependent (Dinev, Goo, Hu, & Nam, 2009), it would be interesting to establish the extent to which these results generalize to other cultural contexts with different approaches to privacy. Furthermore, the sample size allowed detection of only medium sized or bigger differences. Replicating our findings on larger samples will ensure confident detection of smaller differences as well as greater generalizability of our findings.

Furthermore, we considered only two relevant privacy threats (Earp et al., 2005; Krasnova et al., 2009; Milne et al., 2006; Van Slyke et al., 2006). The privacy policy was thus oversimplified and related only to SNSs. It would be interesting to extend the inquiry to additional threats as well as other digital applications, especially in the Internet of Things domain. We argue that it will take substantial effort by the community to comprehensively present all relevant information and yet keep the policies short, simple and so engaging that users truly understand what they consent to.

## 5.2 Conclusion

Informed consent is the pinnacle of user self-protection. Policy makers and scholars have therefore urged platforms to provide users with simple and prominent privacy policies and move beyond the so-called clickwrap ("I agree") contracts of adhesion (MacLean, 2016). However, our study shows that prominent and transparent privacy policies, and especially the threats of secondary PD use, are still miscomprehended by around half of the users. Moreover, in the case of threats of secondary PD use, miscomprehension is more likely to go in the privacy-unsafe direction, meaning users assume privacy-friendlier policies than what is actually presented and consequently they disclose data unwittingly. We also show that beyond self-reported intentions, the threat of secondary PD use might be discarded, as users still upload PD even when the privacy-invasive policies are correctly comprehended. This calls for design solutions that highlight and simplify the threat of secondary PD use, user-friendly options to control such use as well as stricter regulation for secondary PD use and privacy in general.

# Referencess